\documentclass[12pt,a4paper]{article}

\begin{document}

\def\be{\begin{equation}}
\def\ee{\end{equation}}
\def\tr{{\rm tr}}
\def\nn{\nonumber}
\def\g5{{\gamma}^{5}}
\def\ga{{\gamma}}
\def\Ga{{\Gamma}}
\def\Z{{Z\!\!\!Z}}
\def\sm{$ {\rm T}_{\rm SM} $}
\def\bea{\begin{eqnarray}}
\def\eea{\end{eqnarray}}
\def\ba{\begin{array}}
\def\ea{\end{array}}
\def\pa{{\partial}}


\begin{titlepage}
\begin{flushright}   IFT-25/99  \end{flushright}
\begin{flushright}   hep-th/9910161  \end{flushright}

\vspace{1cm}

\begin{center}
{\LARGE{\bf {Towards supersymmetric cosmology}}} 
\end{center}

\begin{center}
{\LARGE{\bf{in M--theory}}}
\end{center}
\vskip .5cm

\vspace{1cm}

\centerline{\bf Krzysztof A. Meissner and Marek Olechowski}

\vskip .5cm
\centerline{\em Institute of Theoretical Physics}
\centerline{\em Warsaw University}
\centerline{\em Ho\.za 69, 00-681 Warsaw, Poland}

\thispagestyle{empty}

\vspace{2cm}

\begin{abstract}

We present a new solution in the heterotic M--theory in which the
metric depends on (cosmic) time. The solution preserves $N=1$
supersymmetry in 4 dimensions in the leading order of the
$\kappa^{2/3}$ expansion. It is the first example of the
time--dependent supersymmetric solution in M--theory on $S^1/\Z_2$. 
It describes expanding 4--dimensional space--time with shrinking
orientifold interval and static Calabi--Yau internal space. 

\end{abstract}

\vfill
\noindent
October 1999

\end{titlepage}


The M--theoretic extension of the heterotic $E_8\times E_8$ string 
leads to a geometric picture of two 10--dimensional walls (branes) at
the ends of a finite interval along eleventh dimension
\cite{HoravaWitten,Witten}.  
While the supergravity multiplet (gravition, gravitino and
antisymmetric tensor fields) can penetrate in the $d=11$ bulk, the
two $E_8$ gauge multiplets are confined to the two walls,
respectively.
The role of the walls can in general be played by
higher dimensional p-branes that support gauge groups. 
When six of the dimensions are compactified (at the lowest
order on a Calabi--Yau space) one obtains phenomenologically promising
models with gauge fields living on the $d=4$ walls. 
One should however remember that such a compactification should be
consistent with the 11--dimensional equations of motion.

Within this heterotic M--theory it is possible to unify the
gravitational constant with the gauge coupling constants at the
M--theory fundamental scale. Assuming such a unification and using
measured values of the Newton's constant $G_N$ and of the gauge
coupling
constants one can deduce scales of compactification. These scales of
the internal dimensions are of the order of $10^{15}$ GeV for the
$x^{11}$ interval and $10^{16}$ GeV for the Calabi--Yau space
\cite{Witten,BD,NOY}.  
These energy scales are very similar to the characteristic scales in
the Early Universe, for example during inflationary period. Therefore
it is natural to ask whether cosmological solutions (with the metric
depending on time) are possible in M--theory.

The problem of finding cosmological solutions in M--theory has been
addressed in several papers. Some of them discuss solutions within
strongly
coupled limit of the heterotic string theory (heterotic M--theory). 
One of the approaches in this case consists in using an effective
5--dimensional theory with walls \cite{LOW}. The domain wall solutions 
\cite{LOSW}
are generalized by introducing time dependence of their parametrs. 
This aproach can be justified \cite{d5} because the scale of the
$x^{11}$ interval is lower than the Calabi--Yau scale. But one should
check {\it a posteriori} whether the 11--dimensional equations of
motion are satisfied and it is a very stringent requirement. In
another
approach one looks for solutions in the theory with walls by
continuation of p--branes to the Euclidean space \cite{Euc}.

Other papers discuss solutions within 11-- or 10--dimensional
supergravities without walls (general M--theory, less promising
phenomenologically). There one assumes different compact, internal
manifolds (for example squashed spheres \cite{squashed} or tori
\cite{eldimsugra}) or compactification by the Scherk--Schwarz
mechanism \cite{scherk} (giving a static Universe with
domain--walls).  

In the case of the heterotic M--theory none of the cosmological
solutions proposed up to now preserves $N=1$ $d=4$ supersymmetry. 
The purpose of this paper is to generalize the result of ref.\
\cite{Witten} and prove that such a supersymmetric solution is
possible.

The low energy limit of the heterotic M--theory is given  
by the following lagrangian \cite{HoravaWitten}:  
\bea
{\cal L}&=& \frac{1}{\kappa^2}\int_{M^{11}}d^{11}x\sqrt g
\left[
-\frac{1}{2}R-\frac{1}{2}\bar\psi_I\Gamma^{IJK}D_J\left(\frac{\Omega+
\hat\Omega}{2}\right)\psi_K-\frac{1}{48}
G_{IJKL}G^{IJKL}\right.
\nn\\
&& \quad -\frac{\sqrt
2}{384}\left(\overline\psi_I\Gamma^{IJKLMN}\psi_N
+12\overline\psi^J\Gamma^{KL}\psi^M\right)\left(G_{JKLM}
+\hat G_{JKLM}\right)\nn\\
&& \quad\quad \left.{}-\frac{\sqrt 2}{3456}\epsilon^{I_1I_2\ldots
I_{11}}
C_{I_1I_2I_3} G_{I_4\ldots I_7} G_{I_8\ldots
I_{11}}
\right]
\label{lagr}
\\
&+&
\frac{1}{4\pi(4\pi\kappa^2)^{2/3}}
\sum_{n=1}^2
\int_{M^{10}_n}d^{10}x\sqrt g
\left[
-\frac{1}{4}F^a_{nXY}F^{a\,XY}_n-\frac{1}{2}\overline\chi^a_n
\Gamma^X D_X(\hat\Omega)\chi^a_n\right.
\nn\\
&& \quad \left.\qquad{}-\frac{1}{8}\overline\psi_X\Gamma^{YZ}
\Gamma^X\left(F^a_{nYZ}+
\hat F^a_{nYZ}\right)\chi^a_n+\frac{\sqrt 2}{48}
\left(\overline\chi^a_n\Gamma^{XYZ}
\chi^a_n\right)\hat G_{XYZ\,11}
\right]
\nn
\eea
where $I,J,K,\ldots=1,2,\ldots, 11$; $X,Y,Z\ldots=1,2,\ldots, 10$ and  
$n=1,2$ counts the 10--dimensional boundaries (walls) of the
space. The first integral in (\ref{lagr}) describes the supergravity
in the
11--dimensional bulk (metric $g_{IJ}$, three-form $C_{IJK}$ and spin
$3/2$ gravitino $\psi_I$) while the second one describes interactions
of  
the super Yang--Mills fields (gauge field strengths $F_n^a$  and
gauginos $\chi_n^a$) living on two 10--dimensional walls. The field
strength of $C$ is given by $G_{IJKL}=(\pa_I C_{JKL} \pm {\rm
permutations})$ and ${\hat G}$ is the supercovariant field
strength. The signature of the space-time manifold $M^{11}$ is 
$(-+ \dots +)$; gamma matrices obey $\{\Ga_I,\Ga_J\}=2g_{IJ}$ and
$\Ga_{I_1I_2\dots I_n}=(1/n!)\left(\Ga_{I_1}\Ga_{I_2} 
\dots \Ga_{I_n}\pm {\rm permutations}\right)$.

In the above lagrangian only the first two
terms in the long wavelength expansion are kept. They are of relative
order $\kappa^{2/3}$. All higher order terms (order $\kappa^{4/3}$ or
higher) will be consistently dropped in this paper. 

We now try to find a time--dependent solution that preserves at least
some of the supersymmetries generalizing the method employed in
\cite{Witten}.   

The supersymmetry transformation law for $\psi_I$ is
\be
\delta\psi_I
=
D_I\epsilon + \frac{\sqrt{2}}{288} (\Ga_{IJKLM}
-8g_{IJ}\Ga_{KLM}){\hat G}^{JKLM}\epsilon
\,.
\label{eq:trans-psi}
\ee
The condition for a spinor $\epsilon$ to generate an unbroken
supersymmetry is the vanishing of the r.h.s.\ of (\ref{eq:trans-psi}): 
\be
D_I\epsilon+ {\sqrt 2\over 288} (\Ga_{IJKLM}
-8g_{IJ}\Ga_{KLM})G^{JKLM}\epsilon=0
\,.
\label{eq:susy}
\ee
In the above equation we use the covariant (and not supercovariant) 
field strength because we assume that vacuum expectation values of all
fermion fields vanish.

To the zeroth order in $\kappa^{2/3}$ the 11--dimensional space--time
in the heterotic M--theory  is of the form 
$M^{11}=M^4 \times X^6 \times S^1/\Z_2$.
Indices $A,B,C,\ldots$ run from 5 to 10 and are tangent to $X^6$;
$a,b,c,\ldots =1,2,3$ are holomorphic indices tangent to $X^6$, and
$\bar a, \bar b, \bar c,\ldots =1,2,3$ are analogous antiholomorphic
indices; $\mu,\nu,\rho,\,\ldots = 1,2,3,4$ are $d=4$ space--time
indices while $i,j,k,\ldots = 2,3,4$ are space indices.

We are looking for a solution of the supersymmetry condition 
(\ref{eq:susy}) in the presence of nonvanishing vev of $G_{JKLM}$ of
the order of $\kappa^{2/3}$. First we perturb the metric: 
\be
ds^2=
\left(\eta_{\mu\nu}+b_{\mu\nu}\right)dx^\mu dx^\nu
+2(g_{a\bar b}+h_{a\bar b})dx^adx^{\bar b}
+(1+c)(dx^{11})^2
\label{eq:ds2}
\ee
where $b$, $h$, and $c$ are functions of order $\kappa^{2/3}$ and we
will neglect higher order corrections. We take the zeroth order metric
on $M^4$ to be the Minkowski one $\eta_{\mu\nu} dx^\mu dx^\nu$.  
$X^6$ is to zeroth order a Calabi-Yau manifold with a metric 
$g_{AB}$ with non-zero components $g_{a\bar b}=g_{\bar b a}$.  
A K\"ahler form on $X^6$ is $\omega_{AB}$ with non-zero components
$\omega_{a\bar b}=-i g_{a\bar b}=-\omega_{\bar b a}$.

Under a perturbation of the metric $g_{IJ}\to g_{IJ}+\delta g_{IJ}$,
to first order in $\delta g$, the covariant derivative of a spinor
changes by $D_I\epsilon \to D_I\epsilon- \frac14 D_J \delta
g_{KI}\Ga^{JK}\epsilon$. Introducing the covariantly constant (with
respect to the unperturbed metric) spinor field $\epsilon_0$ we write 
\be
\epsilon= e^{p} \epsilon_0,
\ee
$p$ being also of order $\kappa^{2/3}$.

Then the first term in eq.\ (\ref{eq:susy}) is
\bea
dx^ID_I\epsilon=
\frac14
&&\!\!\!\!\!\!\!\!\!
\left\{
4dx^I\pa_I p 
+dx^\mu
\left( 
\pa_{11}b_{\mu\rho}\Ga^\rho + \pa_{\bar b}b_{\mu\rho}\Ga^{\rho\bar b} 
-D_\rho b_{\sigma\mu}\Ga^{\rho\sigma}
\right) 
\right.
\nn\\
&&\!\!\!\!\!
\left.  
+dx^a
\left(
-g^{b\bar b}D_b h_{a \bar b} + \pa_{11}h_{a \bar b}\Ga^{\bar b} 
-D_{\bar b}h_{a \bar c}\Ga^{\bar b \bar c} 
-\pa_\mu h_{a \bar b} \Ga^{\mu \bar b}
\right)
\right.
\nn\\
&&\!\!\!\!\!
\left. 
+dx^{\bar b}
\left(
g^{a \bar c}D_{\bar c}h_{a \bar b}\right)
+dx^{11}
\left( -\pa_{\bar b} c \Ga^{\bar b} - \pa_\mu c \Ga^\mu \right)
\right\}
\epsilon
\,.
\label{eq:Deta}
\eea
The $\Z_2$ invariance requires that the supersymmetry transformation
parameter $\epsilon$ obeys chirality condition
$\Ga^{11}\epsilon=\epsilon$. To simplify calculations it is also
convenient to assume that $\Ga^a\epsilon=0$. For spinors which are
singlets of the $SU(3)$ holonomy group of $X^6$ this reduces to the
Weyl condition $\ga^5\epsilon=-\epsilon$ in $d=4$ (where
$\ga^5=-i\Ga^1\Ga^2\Ga^3\Ga^4$ and condition $\Ga^{\bar a}\epsilon=0$
would correspond to the opposite 4--dimensional chirality). From
Weyl--type solutions of the supersymmetry condition we can construct
Majorana--type solutions if needed.

Supersymmetry condition (\ref{eq:susy}) with the perturbed metric 
(\ref{eq:ds2}) can be solved only if some of the components of the 
field strength tensor $G$ are nonzero. It was shown in 
\cite{Witten} that the following components of $G$
\bea
\alpha & = & \omega^{AB} \omega^{CD} G_{ABCD}
\,,
\label{eq:alpha}
\\
\beta_A & = & \omega^{BC} G_{ABC\,11}
\,,
\label{eq:beta}
\\
\theta_{AB} & = & \omega^{CD} G_{ABCD}
\label{eq:theta}
\eea
must be nonzero in the presence of sources. Looking for solutions  
depending on space--time coordinates we have to assume that there are 
more nonvanishing components of $G$. If we are looking for
supersymmetric solutions the only additional components allowed are of
the type 
$G_{\mu 11 a \bar b}$. Therefore, we define
\be
\zeta_\mu = i \omega^{AB} G_{\mu 11 A B}
\,.
\label{eq:zeta}
\ee
There are no sources entering the Bianchi identity for this
quantity. It was argued in \cite{Witten} that "sourceless" components
of $G$ should vanish since $G$ with indices tangent to the internal
space must be cohomologically trivial. This argument does not apply to
$G_{\mu 11 a \bar b}$ because $d=4$ space--time is not compact and
boundaryless so even without sources allows for constant field
strengths.

With the above definitions the second term in the supersymmetry
condition (\ref{eq:susy}) can be written as 
\bea
\frac{\sqrt 2\,dx^I}{ 288 }
&&\!\!\!\!\!\!\!\!
\left(\Ga_{IJKLM}-8g_{IJ}\Ga_{KLM}\right)
G^{JKLM}\epsilon
=
\nn\\
&&\!\!\!\!\!\!\!\!
= \frac{\sqrt 2}{288} 
\left\{
dx^\mu 
\left(
   24 \zeta_\mu -3\alpha\Ga_\mu - 
   12i\eta_{\mu\rho}\beta_{\bar b}\Ga^{\rho\bar b}
   -12 \eta_{\mu\rho}\zeta_\sigma\Ga^{\rho\sigma} 
\right)
\right.
\nn\\
&&\,\,\,\,\,\,\,\,\,\,\,\,
+dx^a
\left(
36 i \beta_a
+\left(36 i \theta_{a\bar b}-3\alpha g_{a\bar b}\right)\Ga^{\bar b}
+\left(36G_{11 a\bar b\bar c}+12ig_{a\bar c}\beta_{\bar b}\right) 
       \Ga^{\bar b\bar c}
\right.
\nn\\
&&
\left.\,\,\,\,\,\,\,\,\,\,\,\,\,\,\,\,\,\,\,\,\,\,\,\,\,\,\,\,\,
+\left(72 G_{\mu 11 a {\bar b}} + 12 g_{a {\bar b}}\zeta_\mu \right)
\Ga^{\mu {\bar b}}
\right)
\nn\\
&&\,\,\,\,\,\,\,\,\,\,\,\,
\left.
+dx^{\bar b} \left( 12 i \beta_{\bar b} \right)
+dx^{11}
\left(-3\alpha -24i\beta_{\bar b}\Ga^{\bar b}
-24\zeta_\mu\Ga^\mu\right)
\right\}\epsilon
\,.
\label{eq:GaG}
\eea

Substituting (\ref{eq:Deta}) and (\ref{eq:GaG}) into (\ref{eq:susy}) 
we obtain the following equations for the perturbations of the metric
and the supersymmetry generating spinor:
\bea
D_\rho b_{\sigma\mu} - D_\sigma b_{\rho\mu}
&=& \frac{\sqrt{2}}{6}
\left(g_{\mu\sigma}\zeta_\rho-g_{\mu\rho}\zeta_\sigma\right)
\,,
\label{eq:b}
\\
\pa_\mu h_{a \bar b} &=& 
\frac{\sqrt{2}}{6} g_{a \bar b}\zeta_\mu + \sqrt{2} G_{\mu 11 a \bar
b}
\,,
\label{eq:h}
\\
\pa_\mu c &=& - \frac{\sqrt{2}}{3} \zeta_\mu
\,,
\label{eq:c}
\\
\pa_\mu p &=& - \frac{\sqrt{2}}{12} \zeta_\mu
\,,
\label{eq:p}
\\
\pa_{\bar a} b_{\mu\nu}&=&\frac{\sqrt{2}}{6}i\beta_{\bar
a}\eta_{\mu\nu}
\,,
\label{eq:ba}
\\
\pa_{11} b_{\mu\nu}&=&\frac{\sqrt{2}}{24}\alpha \eta_{\mu\nu}
\,.
\label{eq:b11}
\eea
The other equations are the same as those obtained in \cite{Witten}.

One can check that eq.\ (\ref{eq:susy}) has no solutions with nonzero
components of $G$ other than those given in 
eqs.\ (\ref{eq:alpha}--\ref{eq:zeta}).

In order to check the consistency of the above set of equations we use
the following Bianchi identity
\be
(dG)_{\mu \nu 11 a \bar b}
=\pa_\mu G_{\nu 11 a \bar b} - \pa_\nu G_{\mu 11 a \bar b}
=0
\,.
\label{eq:B1}
\ee
Contracting it with $\omega^{a \bar b}$ we get
\be
\pa_\mu \zeta_\nu - \pa_\nu \zeta_\mu = 0
\,.
\label{eq:B2}
\ee
It is easy to check that equations (\ref{eq:B1}) and (\ref{eq:B2}) are
the integrability conditions for equations (\ref{eq:b})--(\ref{eq:p})
(integrability conditions for (\ref{eq:ba}) and (\ref{eq:b11}) were
discussed in \cite{Witten}).

In the potential application of the solutions of
(\ref{eq:b})--(\ref{eq:b11}) to cosmological
situations the natural assumption is to allow for dependence of all
fields on the cosmic time ($t=x^1$) but not on the space coordinates
(we are perturbing flat Minkowski space--time so spaces with
topologically different space-like sections are excluded). Therefore
we assume 
\be
G_{i 11 a \bar b}=0
\,.
\ee
The equation of motion for $G$ reads
\bea
D_I G^{I J K L} 
&\!\!\!=&\!\!\! 
\frac{\sqrt{2}}{1152}
\epsilon^{J K L I_4  \ldots I_{11}}
G_{I_4 I_5 I_6 I_7}G_{I_8 I_9 I_{10} I_{11}}
\nn\\
&\!\!\!-&\!\!\!
\frac{1}{1728\pi}\left(\frac{\kappa}{4\pi}\right)^{2/3}
\delta\left(x^{11}\right)
\epsilon^{11 J K L I_5 \ldots I_{11}}
{\rm Tr}\left(AdA+\frac23 A^3\right)_{I_5 I_6 I_7}G_{I_8 I_9 I_{10}
I_{11}}
.
\nn\\
\label{eq:motion}
\eea
The r.h.s.\ of the above equation may be nonzero for general vacuum 
expectation values of $G$ given by eqs.\
(\ref{eq:alpha}--\ref{eq:zeta}). However, as shown in ref.\
\cite{Witten}, quantities $\alpha$, $\beta_A$ and $\theta_{AB}$ are
all of order 
$\kappa^{2/3}$. We are going to restrict our analysis to situations
when $\zeta_\mu$ is also of the order of $\kappa^{2/3}$ ($\zeta_\mu$
of order 1 would not satisfy equations of motion at the leading
order). For $G$ satisfying this condition the r.h.s. of eq.\
(\ref{eq:motion}) is of higher order ($\kappa^{4/3}$) and should be
dropped.

{}From now on we will use for simplicity 
the cartesian coordinates to parametrize $M^4$.
Then, the equations of motion and Bianchi identities relevant for us
are:
\be
\pa_{11}G_{t 11   a \bar b} = 0
\,,
\ee
\be
- \pa_t G_{t 11 a \bar b} + D^c G_{c 11 a \bar b} 
+ D^{\bar c} G_{\bar c  11 a \bar b} = 0
\,,
\ee
\be
(DG)_{t 11 a b \bar c}=(DG)_{t 11 a \bar b \bar c} = 
(DG)_{t i 11 a \bar b}=0
\,.
\ee

By introducing $G_{t 11 a \bar b}$ we don't want to change to the 
lowest order the $G$ components in the  solution of \cite{Witten} 
so we assume
\be
\pa_t G_{t 11 a \bar b}=\pa_i G_{t 11 a \bar b}=\pa_c G_{t 11 a \bar
b}=
\pa_{\bar c} G_{t 11 a \bar b}=0
\,.
\label{eq:dGzero}
\ee
We can do this in a consistent way because the Bianchi identities for
the added components of $G$ ($G_{t 11 a \bar b}$) have no sources at
the walls.

Contracting (\ref{eq:dGzero}) with $\omega^{a \bar b}$ we get
\be
\zeta_1 = {\rm const}
\,.
\label{eq:zeta0}
\ee

The explicit solution to eqs.\ (\ref{eq:b})--(\ref{eq:b11}) reads (we
drop $x^a$ and $x^{\bar a}$ dependence analyzed in \cite{Witten}):
\bea
b_{tt}(t,x^{11}) 
&\!\!\!=&\!\!\! 
\left(\bar\alpha x^{11} 
+ f_1(t)\right)\eta_{tt}
\,,
\\
b_{ij} (t,x^{11})
&\!\!\!=&\!\!\! 
\left(\bar\alpha x^{11} + \zeta t\right)\eta_{ij}
\,,
\\
c(t,x^{11})
&\!\!\!=&\!\!\! 
-2\left(f_2(x^{11}) + \zeta t\right)
\,,
\\
p(t,x^{11}) 
&\!\!\!=&\!\!\! 
-\frac12\left(\bar\alpha x^{11} + \zeta t\right)
\,.
\eea
where $\zeta=\frac{\sqrt{2}}{6}\zeta_1$, $\,\bar\alpha$ is an average
value over the Calabi--Yau space of $\frac{\sqrt{2}}{24}\alpha$
and $f_1(t)$ and $f_2(x^{11})$ are arbitrary functions reflecting
the freedom of coordinate choice of time and $x^{11}$. 
We have dropped the arbitrary constants 
to make the result look simpler.

It is important to analyse the volume of $X^6$ because it determines
the value of the gauge coupling constants. Its dependence on the
space--time coordinates is given by
\be
\pa_\mu V_6 
=
\pa_\mu \int d^6x \sqrt{g^{(6)}}
=
\int d^6x \sqrt{g^{(6)}} g^{a \bar b} \pa_\mu h_{a \bar b}
\,.
\ee
Substituting (\ref{eq:h}) into this formula we get
\bea
\pa_\mu V_6 
&\!\!\!=&\!\!\!
\int d^6x \sqrt{g^{(6)}} g^{a \bar b} 
\left(\frac{\sqrt{2}}{6} g_{a \bar b} \zeta_\mu 
     + \sqrt{2} G_{\mu 11 a \bar b} \right) 
\nn\\
&\!\!\!=&\!\!\! 
\int d^6x \sqrt{g^{(6)}} 
\left(\frac{\sqrt{2}}{2} \zeta_\mu 
     + \sqrt{2} g^{a \bar b} G_{\mu 11 a \bar b} \right) 
= 0
\,.
\label{eq:V6}
\eea
In these equalities we used the definition of $\zeta_\mu$ 
(eq.\ (\ref{eq:zeta})) and the fact that the complex dimension 
of $X^6$ is equal to 3. Thus, $V_6$ depends only on $x^{11}$ 
but not on time:
\be
V_6(x^{11}) 
= 
V_6(0)
-\frac{\sqrt{2}}{8} x^{11}\int d^6 x \sqrt{g^{(6)}}
\,\alpha
\,.
\label{eq:v6}
\ee

This fact that the volume of Calabi--Yau space, $V_6$, does not depend
on time (as shown in eq.\ (\ref{eq:V6})) has important consequences. 
The squares of the gauge coupling constants are inversely proportional
to the volume of the Calabi--Yau space $V_6$. Therefore the gauge
coupling constants do not depend explicitly on time.

On the other hand, the scale factor of the 3--dimensional space and 
the length of $S^1/\Z_2$ do change in time. 
To discuss possible expansion or contraction of internal or space-time
dimensions it is convenient to fix $f_1(t)$. Apart from the $x^{11}$
dependence, the usual cosmic time prescription amounts to putting
$f_1(t)=0$  
and that is our choice (we also fix the $x^{11}$ coordinate freedom by
choosing $f_2(x^{11})=0$).
It is interesting that
when we choose the sign of $\zeta$ positive so that it describes
expanding space (space Hubble parameter positive) then the solution
describes shrinking $x^{11}$ scale factor.

Since the length of the $x^{11}$ interval in this solution may depend
on time we have to reconsider the question of the connection between
the Newton's constant and the gauge coupling constant at the $GUT$
scale. 
For the time--independent solutions these relations read 
\be
G_N 
= \frac{\kappa_4^2}{8 \pi} 
= \frac{\kappa^2}{8 \pi R_{11} {\bar V}_6}
\,,
\label{eq:GN}
\ee
\be
\alpha_{GUT} = \frac{(4\pi\kappa^2)^{2/3}}{V_6^0}
\label{eq:alphaGUT}
\ee
where $R_{11}=\pi\rho$ in the length of the $S^1/\Z_2$ interval,
$V_6^0=V_6(x^{11}=0)$ is the volume of $X^6$ at our wall and
${\bar V}_6$ is the averaged (over $x^{11}$) volume of
$X^6$. 
Eliminating $\kappa$ we get 
\be
R_{11}  = \frac{\left(\alpha_{GUT} V_6^0\right)^{3/2}}{32\pi^2 G_N
{\bar V}_6}
\,.
\label{eq:alGUT}
\ee
Since in the linear approximation
\be
\frac12 V_6^0\le {\bar V}_6\le V_6^0 
\ee
then knowing $\alpha_{GUT}$ and $G_N$ and assuming 
$V_6^0\approx M_{GUT}^{-6}$ we can calculate $R_{11}$ -- it turns out
to be of the order
\be
R_{11}^{-1}\approx 10^{15} {\rm GeV}
\,.
\label{eq:R11}
\ee
Plugging this value back into eq.\ (\ref{eq:v6}) we have to require
that the volume of $X^6$ at the other wall is positive which gives an
upper bound on the length of $S_1/\Z_2$: $R_{11}<R_{11}^{crit}$. This
critical length $R_{11}^{crit}$ turns out to be close to the value
(\ref{eq:R11}) obtained from arguments based on phenomenology. 
One could expect that the evolution of the length of the $x^{11}$
interval could violate the condition $R_{11}<R_{11}^{crit}$ 
for large enough negative $t$. However, this is not the case.  
$R_{11}^{crit}$ is obtained from the condition that $V_6$ must be
positive at both walls. We can see from eq.\ (\ref{eq:v6}) that since
$x^{11}$ is the ``comoving'' coordinate the volume $V_6$ does not
change with time and stays positive along the whole $x^{11}$
interval. The physical length of the interval changes with time but
this is ``compensated'' by the change of $R_{11}^{crit}$ induced by
the change of the Newton's constant as shown in eq.\ (\ref{eq:GN}).

Let us analyse the resulting metric in some more detail. We assume (as
everywhere else in the paper) that we keep only linear terms in
$\kappa^{2/3}$. Therefore the 5-dimensional metric reads
\be
ds^2=-\left(1+\bar\alpha x^{11}\right)dt^2+
\left(1+\bar\alpha x^{11}+\zeta t\right)(dx^i)^2
+(1-2\zeta t)(dx^{11})^2
\,.
\label{eq:dscosm}
\ee
The Riemann tensor calculated from the above metric has
an expansion in powers of $\bar\alpha$ and $\zeta$ starting from
quadratic terms. Therefore the Riemann tensor components
are of the order $\kappa^{4/3}$ and the equations of motion relating
the Einstein tensor with the square of $G$ are satisfied to order
$\kappa^{2/3}$ (it is a similar situation as in \cite{Witten}).

The metric (\ref{eq:dscosm}) may depend on time while preserving
4 supersymmetry charges corresponding to $N=1$ supersymmetry in
$d=4$ (this is similar to the anti de Sitter space where
supersymmetry can be preserved although the
metric depends on the spatial coordinates).  
This fact is extremely important since preserved supersymmetry 
in a given background (with globally well defined fields) guarantees
that this background satisfies the 11--dimensional equations of
motion. Thus, the solution described in this paper is the first
approximation (in the $\kappa^{2/3}$ expansion) to the time--dependent
solutions in M--theory and it is the first example of such a solution.

Since 4--dimensional supersymmetry is preserved one should be able to
identify an appropriate 4--dimensional gravitino field. This could be
done using methods of ref.\ \cite{MNO} generalized to the time
dependent case.

The metric (\ref{eq:dscosm}) is known in the linear approximation
only, so we cannot make any definite statements about the global
behaviour of the solution with time i.e. whether it is an
inflationary-like solution or a power-like one or anything else. 
A similar problem exists already in the case of $x^{11}$
dependence. There we also know the metric only in the linear
approximation, but the problem with the global behaviour is less
severe because the length of $x^{11}$ interval is finite (although
large).

One may note that the presence of nonzero $G_{t 11 a \bar b}$ may be
important for the cosmological constant problem. It stems from the
fact that the contribution of these components to vacuum energy has
the opposite sign compared to the contribution of other components of
$G$
present in the static solution.

If we extend validity of the metric (\ref{eq:dscosm}) to arbitrary
time we can derive a rather suprising lower bound on the value of
$G_{t 11 a \bar b}$. The argument goes as follows. One can see that
the speed of light in the $x^{11}$ direction is given by  
\be
v_{11}^2=\frac{1+\bar\alpha x^{11}}{1-2\zeta t}
\,.
\ee
We can also introduce a velocity with which a fixed space scale factor
``travels'' along the $x^{11}$ direction:
\be
v_{sf}=-\frac{\zeta}{\bar\alpha}
\,.
\ee
The metric (\ref{eq:dscosm}) describes an initial singularity (space
scale factor vanishes) for 
\be
\bar\alpha x^{11}+\zeta t=-1
\ee
so the singularity at the other wall ($x^{11}=\pi\rho$) is earlier
than
at our wall ($x^{11}=0$). On physical grounds the signal about the
beginning of expansion from a
point on $x^{11}$ should not reach a nearby point
before the expansion has started there.
Therefore the speed of light $|v_{11}|$ should not be greater
than $|v_{sf}|$ at each point on $x^{11}$ axis. This condition gives
\be
\zeta\ge \bar\alpha\,\sqrt
{\frac{1+\bar\alpha x^{11}}{3+2\bar\alpha x^{11}}}\ge
\frac{\bar\alpha}{\sqrt{3}}
\,.
\ee
In terms of our original fields it can be translated as
\be
\int_{X^6}G_{t\,11\,AB}\omega^{AB}
> \frac{4}{\sqrt{3}}\,\int_{X^6}G_{ABCD}\omega^{AB}\omega^{CD}
\,.
\ee
This lower bound on $\int_{X^6}G_{t\,11\,AB}\omega^{AB}$ is quite
surprising. It shows that whenever we switch on nonzero vev for
$G_{ab\bar a \bar b}$ (what we anyway should do due to sources on the
walls
in the Bianchi identities) while keeping supersymmetry then the
solution
is physically unstable unless we switch on also sufficiently large vev
for $G_{t 11 a \bar b}$. Then the solution is necessarily
time--dependent. One should remember however, that the bound can be
modified by presently unknown higher order (in $\kappa^{2/3}$) terms.

In conclusion, we have presented the first example of the time
dependent solution in the heterotic M--theory that preserves $N=1$
$d=4$ supersymmetry. 
It describes expanding 4--dimensional space--time with shrinking
orientifold interval and static Calabi--Yau internal space. 
As a result the gauge coupling constants do not depend on time while
the Newton's constant increases with time. 
It remains to be seen what are the global properties of such
cosmological solutions in M--theory and how they can be used to
describe the evolution of the early Universe.

\vspace{2cm}
\noindent
{\Large \bf Acknowledgements}
\vskip .5cm
K.A.M. was partially supported by the Polish KBN grant 2P03B 03715
(1998-2000) and by the Polish--French Cooperation Grant POLONIUM 1999.
M.O. was partially supported by the Polish grant 
KBN 2 P03B 052 16 (1999-2000).


\newpage


\begin{thebibliography}{Ref}

\bibitem{HoravaWitten} 
P. Ho\v{r}ava and E. Witten, 
Nucl. Phys. B460 (1996) 506; B475 (1996) 94.

\bibitem{Witten} 
E. Witten, Nucl. Phys. B471 (1996) 135.

\bibitem{BD} 
T. Banks and M. Dine, Nucl. Phys. B479 (1996) 173.

\bibitem{NOY} 
H.P. Nilles, M. Olechowski and M. Yamaguchi,
Phys. Lett. B415 (1997) 24; Nucl. Phys. B530 (1998) 43.

\bibitem{LOW} 
A. Lukas, B.A. Ovrut, and D. Waldram, 
Nucl. Phys. B495 (1997) 365; 
Nucl. Phys. B509 (1998) 169;
Phys. Rev. D60:86001 (1999);
A. Lukas, B.A. Ovrut, Phys. Lett. B437 (1998) 291; 
H.S. Reall, Phys. Rev. D59:103506 (1999).

\bibitem{LOSW} 
A. Lukas, B.A. Ovrut, K.S. Stelle and D. Waldram, 
Phys. Rev. D59:86001 (1999).

\bibitem{d5}
A. Lukas, B.A. Ovrut, K.S. Stelle and D. Waldram, 
Nucl. Phys. B552 (1999) 246;
J. Ellis. Z. Lalak, S. Pokorski and S. Thomas, hep--th/9906148. 


\bibitem{Euc} 
K. Benakli, Int. J. Mod. Phys. D8 (1999) 153; 
Phys. Lett. B447 (1999) 51.  

\bibitem{squashed} 
M. Bremer, M.J. Duff, H. Lu, C.N. Pope, and K.S. Stelle, 
Nucl. Phys. B543 (1999) 321; 
S.W. Hawking and H.S. Reall, Phys. Rev. D59:023502 (1999).

\bibitem{eldimsugra} 
H. Lu, J. Maharana, S. Mukherji, and C.N. Pope, 
Phys. Rev. D57 (1998) 2219; 
H. Lu, S. Mukherji, C. N. Pope, and K.--W. Xu, 
Phys. Rev. D55 (1997) 7926;
H. Lu, S. Mukherji, and C.N. Pope, hep-th/9612224; 
N. Kaloper, I.I. Kogan, and K.A. Olive, 
Phys. Rev. D57 (1998) 7340; Erratum, ibid. D60:049901 (1999). 
A. Feinstein and M.A. Vazquez--Mozo, hep-th/9906006; 
A. Billyard, A. Coley, J. Lidsey and U. Nillson, hep--th/9908102.

\bibitem{scherk}
M.J. Duff, P. Hoxha, H. Lu, R.R. Martinez--Acosta, and C.N. Pope, 
Phys. Lett. B451 (1999) 38; . 

\bibitem{MNO}
K.A. Meissner, H.P. Nilles and M. Olechowski, 
hep--th/9905139, to appear in Nucl. Phys. B.





\end{thebibliography}
\end{document}